\documentstyle[prl,aps,epsfig,amsmath,multicol]{revtex}

\newcommand{\lsim}{\mathrel{\mathop{\kern 0pt \rlap
  {\raise.2ex\hbox{$<$}}}
  \lower.9ex\hbox{\kern-.190em $\sim$}}}
\newcommand{\gsim}{\mathrel{\mathop{\kern 0pt \rlap
  {\raise.2ex\hbox{$>$}}}
  \lower.9ex\hbox{\kern-.190em $\sim$}}}
\newcommand{\one}{\leavevmode\hbox{\small1\normalsize\kern-.33em1}}

\title{Minimum Detection Efficiencies for a Loophole--Free Bell--type Test}

\author{G. Garbarino}

\address{Dipartimento di Fisica Teorica, Universit\`a di Torino
and INFN, Sezione di Torino, I--10125 Torino, Italy}

\date{\today}

\begin{document}
\draft
\maketitle

\begin{abstract}
We discuss the problem of finding the most favorable conditions 
for closing the detection loophole in a test of 
local realism with a Bell inequality. 
For a generic non--maximally entangled 
two--qubit state and two alternative measurement bases we apply
Hardy's proof of non--locality without inequality 
and derive an Eberhard--like inequality.
For an infinity of non--maximally entangled states we find that 
it is possible to refute local realism by requiring
perfect detection efficiency for only one of the two measurements:
the test is free from the detection loophole for any value
of the detection efficiency corresponding to the other measurement.
The maximum tolerable noise in a loophole--free test is also
evaluated.
\end{abstract}

\maketitle

\begin{multicols}{2}

According to Bell theorem \cite{Be64}, the quantum--mechanical correlations
shown in an ideal experiment
by the separate parties of an entangled state are so strong that
cannot be reproduced by any local realistic model.

Several experiments with Bell inequalities \cite{Cl69,Cl74,Cl78} 
have been performed to test local realism vs quantum mechanics 
\cite{As82,We98,Ti99,Ro01,Ma08}.
No one of these experiments allowed a conclusive refutation of
local realism, i.e., the violation of a genuine Bell inequality.
Local hidden--variable models exploiting the non--ideal behaviour
of the apparata exist which reproduce the results obtained in each one
of these tests \cite{Cl74,santos-models}.

The reason behind the impossibility of a conclusive Bell--type test
is the persistence of the locality and the detection loopholes. 
For a bipartite system, the locality loophole arises when the  
two (left and right) joint measurements required in the Bell--test
are performed in space--time regions which are not space--like 
separated from each other. In this case, one cannot exclude that
information on the measurement settings
is exchanged between the two measurement regions. 
Besides, when the detection efficiency in a Bell--test is
smaller than a certain critical value, or the test suffers from 
noise beyond a certain threshold, a local realistic model
can be constructed so as to reproduce the quantum--mechanical
correlations entering the Bell inequality. In this case, the test
is affected by the detection loophole, and only 
non--genuine Bell inequalities (incorporating supplementary
assumptions) can be tested experimentally. 

For any bipartite and entangled state one can derive
Bell inequalities without the introduction of (plausible but not testable)
supplementary assumptions concerning undetected events \cite{Cl74,Cl78,Pe70}.
In particular, the most appropriate inequality for confronting 
local realism vs quantum mechanics was
derived long ago by Clauser and Horne~\cite{Cl74}.
For maximally (non--maximally) entangled states, if one assumes
that all the involved measurements are performed with the same
overall detection efficiency $\eta$,
these Clauser--Horne inequalities are violated by quantum mechanics 
only if $\eta >2/(1+\sqrt{2}) \simeq 0.83$ \cite{Ga87} 
($\eta > 2/3\simeq 0.67$ \cite{Eb93}).

Only the recent tests with entangled 
ions of Refs.~\cite{Ro01,Ma08}
closed the detection loophole.
On the contrary, the locality loophole was closed using
entangled photons \cite{We98}. No
experiment closing simultaneously both the detection and the 
locality loopholes has been performed so far.

To find a solution to the detection loophole problem, two
approaches are possible: one can either identify 
apparata and detectors allowing the highest detection 
efficiencies (see for instance the use of homodyne detection 
in continuous--variables Bell--tests \cite{Ga95}) or 
search for new Bell inequalities and/or entangled states
allowing the use of less efficient detectors and sustaining
the maximum amount of noise,
as recently done in Ref.~\cite{La01,Ma03,Ca07,Br07,Ca09}.
Here we further consider the question of performing a Bell--test 
with inefficient experimental apparata.
By using bipartite non--maximally entangled states for which
Hardy's proof of Bell theorem without inequalities applies,
we demonstrate that it is possible to perform a genuine Bell--type 
test by requiring perfect detection efficiency for only one of the two
observables to be alternatively measured on each one of the 
two parties.

We emphasize that, apart form the obvious 
importance for the foundations of quantum mechanics,
the question of a genuine violation of Bell inequalities
is also relevant in connection with quantum information theory.
Indeed, the existence of some
secure quantum key distribution protocols is closely related to 
the loophole--free violation of Bell inequalities \cite{Ba05}.

We start our discussion by determining 
the most general non--maximally entangled state suitable
for proving Hardy's contradiction without inequalities
between local realism and quantum mechanics \cite{Ha93}.
Let us introduce two incompatible qubit bases $\{|a_+\rangle,|a_-\rangle\}$
and $\{|b_+\rangle,|b_-\rangle\}$ of eigenvectors
of the observables $\hat a_\pm=|\pm_a\rangle\langle\pm_a|$
and $\hat b_\pm=|\pm_b\rangle\langle\pm_b|$ with eigenvalues 
$a_\pm=\pm 1$ and $b_\pm=\pm 1$. In general,
the two bases are related to each other by:
\begin{eqnarray}
|+_b\rangle &=& \alpha |+_a\rangle + \beta e^{i\phi}|-_a\rangle~, \\
|-_b\rangle &=& -\beta e^{-i\phi}|+_a\rangle + \alpha |-_a\rangle~,\nonumber
\end{eqnarray}
$\alpha$ and $\beta$ being real numbers with $\alpha^2+\beta^2=1$.

Hardy's proof \cite{Ha93} is applied to a set of four joint probabilities,
three of which are vanishing;
we choose to use the following quantum--mechanical values:
\begin{eqnarray}
\label{prima}
P^{\rm I}_{\rm QM}(a_+,a_+)&\neq&0~, \\
\label{seconda}
P^{\rm I}_{\rm QM}(a_+,b_-)&=&0~, \\
\label{terza}
P^{\rm I}_{\rm QM}(b_-,a_+)&=&0~, \\
\label{quarta}
P^{\rm I}_{\rm QM}(b_+,b_+)&=&0~,
\end{eqnarray}
where the index I reminds us that we are considering
the ideal case of perfect experimental apparata. 

The most general non--maximally entangled state satisfying
the prediction of Eq.~(\ref{quarta}) is:
\begin{equation}
|\psi\rangle=A|+_b\rangle|-_b\rangle
+B|-_b\rangle|+_b\rangle+C|-_b\rangle|-_b\rangle~,
\end{equation}
where $|A|^2+|B|^2+|C|^2=1$.

The vanishing values of the joint probabilities of Eqs.~(\ref{seconda}) 
and (\ref{terza}) require $A\alpha -C\beta e^{-i\phi}=0$ and 
$B\alpha -C\beta e^{-i\phi}=0$, respectively. The solution
of these equations which satisfies the normalization to one
of $|\psi\rangle$ and of the two qubit bases 
$\{|a_+\rangle,|a_-\rangle\}$ and 
$\{|b_+\rangle,|b_-\rangle\}$ is:
\begin{eqnarray}
A=B&=&\sqrt{\frac{1-\alpha^2}{2-\alpha^2}}e^{-i\phi}~, \\
C&=&\frac{\alpha}{\sqrt{2-\alpha^2}}~.
\end{eqnarray}

The state $|\psi\rangle$ for which the three 
quantum--mechanical predictions (\ref{seconda})--(\ref{quarta}) are fulfilled,
which is called Hardy's state, is thus:
\begin{eqnarray}
\label{hardy}
|\psi_H\rangle&=&\frac{1}{\sqrt{2-\alpha^2}}
\left[\sqrt{1-\alpha^2}e^{-i\phi}(|+_b\rangle|-_b\rangle
+|-_b\rangle|+_b\rangle) \right.\\
&&\left. +\alpha|-_b\rangle|-_b\rangle \right]~, \nonumber
\end{eqnarray}
while Hardy's fraction (i.e., the non--vanishing probability
of Hardy's reasoning) turns out to be:
\begin{equation}
\label{prima-2}
P^{\rm I}_{\rm QM}(a_+,a_+)=\frac{(1-\alpha^2)^2\alpha^2}{2-\alpha^2}~,
\end{equation}
and assumes the maximum value
of $(5\sqrt{5}-11)/2\simeq 0.0902$ 
when $\alpha=\alpha_H\equiv \sqrt{(3-\sqrt{5})/2}\simeq 0.618$.

The contradiction without inequalities
between local realism and quantum mechanics applies to the case
of ideal measurements and consist in showing that there is no local 
realistic theory which reproduces the predictions of
Eqs.~(\ref{seconda}), (\ref{terza}), (\ref{quarta}) and (\ref{prima-2}). 
This has been proved in Ref.~\cite{Ha93}.

Moreover, a Bell inequality in the form due to Eberhard \cite{Eb93}
can be deduced which generalizes Hardy's incompatibility proof
to the case of a real test \cite{Ha94}, which has to confirm null events
with imprecise state preparation and measurements. 
Three different Eberhard inequalities correspond 
to the incompatibility proof adopting the joint probabilities
(\ref{seconda}), (\ref{terza}), (\ref{quarta}) and (\ref{prima-2}).
The more convenient inequality for closing the detection loophole turns 
out to be:
\begin{eqnarray}
\label{eberhard}
H_{\rm LR} &\equiv& P(a_+,a_+)/\left[P(a_+,b_-) + P(b_+,b_+) \right. \\
&& \left. + P(b_-,a_+) + P(a_+,b_0)+P(b_0,a_+) \right]\leq 1~, \nonumber
\end{eqnarray}
where the outcome denoted by $b_0$ corresponds to
the cases in which, due to imperfect experimental apparata, 
the measurement in the $\{|b_+\rangle,|b_-\rangle\}$ basis does not
produce an outcome. 

The previous Eberhard inequality can be 
equivalently rewritten in the form of a Clauser--Horne inequality
\cite{Cl74}:
\begin{eqnarray}
\label{c-h}
&&P(a_+,a_+) + P(a_+,b_+) + P(b_+,a_+)-P(b_+,b_+) \\
&&\leq P(a_+,*)+P(*,a_+)~, \nonumber
\end{eqnarray}
with the single--side probabilities given by:
\begin{eqnarray}
P(a_+,*)&=&P(a_+,b_+)+P(a_+,b_-)+P(a_+,b_0)~, \\
P(*,a_+)&=&P(b_+,a_+)+P(b_-,a_+)+P(b_0,a_+)~. \nonumber
\end{eqnarray}

The quantum--mechanical values of the non--vanishing probabilities 
appearing in Eberhard inequality are:
\begin{eqnarray}
P_{\rm QM}(a_+,a_+)&=&\frac{(1-\alpha^2)^2\alpha^2}{2-\alpha^2}\eta_{L,a}\, \eta_{R,a}~,\\
P_{\rm QM}(a_+,b_0)&=& \frac{(1-\alpha^2)^2}{2-\alpha^2} \eta_{L,a}(1-\eta_{R,b})~,\\
P_{\rm QM}(b_0,a_+)&=& \frac{(1-\alpha^2)^2}{2-\alpha^2} \eta_{R,a}(1-\eta_{L,b})~,
\end{eqnarray}
where we have considered different overall detection efficiencies
for the four measurements (two on the left ($L$) and two on the right ($R$)) 
involved in the inequality.

Inequality (\ref{eberhard}) 
(and (\ref{c-h})) is thus violated by quantum mechanics when
the four detection efficiencies of the problem satisfy:
\begin{equation}
\label{viol1}
H_{\rm QM}=\frac{\alpha^2 \eta_{L,a}\eta_{R,a}\,}
{\eta_{L,a}(1-\eta_{R,b})+\eta_{R,a}(1-\eta_{L,b})} >1~.
\end{equation}

Let us considered the following special cases: \\
\emph{Case 1}: $\eta\equiv \eta_{L,a} = \eta_{R,a} = \eta_{L,b} = \eta_{R,b}$,
as for photon--photon \cite{As82,We98,Ti99} and atom--atom \cite{Ro09} entanglement. 
Eq.~(\ref{viol1}) is satisfied when 
\begin{equation}
\eta > 2/(2+\alpha^2)~.
\end{equation}
The minimum value of the detection efficiency, $\eta_{\rm min}=2/3\simeq 0.67$,
is found for $\alpha=1$. Note however that $\alpha$ cannot be identically
equal to 1, otherwise our entangled state would be a factorized one: 
$|\psi_H\rangle\to |-_b\rangle|-_b\rangle= |-_a\rangle|-_a\rangle$,
and all probabilities entering Eberhard inequality would be vanishing.
Note also that the above result for $\eta_{\rm min}$ is analogous to
what found by Eberhard, with a numerical approach, in Ref.~\cite{Eb93}.

\emph{Case 2}: left--right asymmetric measurements, 
$\eta_L\equiv \eta_{L,a} = \eta_{L,b}$ and $\eta_R\equiv \eta_{R,a} = \eta_{R,b}$. 
This is the case of atom--photon entanglement \cite{Vo06},
for which the measurements on the atom can be done with high
efficiencies. For a given efficiency $\eta_R$, Eq.~(\ref{viol1}) is satisfied when
\begin{equation}
\eta_L > \eta_R/[(\alpha^2+2)\eta_R-1]~,
\end{equation}
and again $\eta_L$ is minimized for $\alpha=1$.
Let us consider the particular case in which measurements on the right
party are done with 100\% efficiency. One has:
\begin{equation}
\eta_L > 1/(\alpha^2+1) \,\,\,\,\,\, {\rm when} \,\,\,\,\,\, \eta_R=1~.    
\end{equation}
A result analogous to our one:
$\eta_L>1/2$ when $\eta_R=1$ and $\alpha=1$,
has been obtained in Ref.~\cite{Ca07}, but starting
from a different bipartite non--maximally entangled state and
by using four different measurement bases (two 
for each one of the two parties, as in standard Bell--tests).

\emph{Case 3}: observable asymmetric measurements,
$\eta_a\equiv \eta_{L,a} = \eta_{R,a}$ and $\eta_b\equiv \eta_{L,b} = \eta_{R,b}$.
This is the case, for instance, of entangled neutral kaons \cite{Br06},
for which lifetime is measurable with a much larger efficiency than strangeness.
For a given efficiency $\eta_b$, Eq.~(\ref{viol1}) is satisfied when
\begin{equation}
\eta_a > 2(1-\eta_b)/\alpha^2~.
\end{equation}
When the measurement in the $\{|b_+\rangle,|b_-\rangle\}$ basis are possible 
with 100\% efficiency, Eberhard inequality is violated by
quantum mechanics independently of the value of $\eta_a$: 
\begin{equation}
\label{best}
{\rm Violation\,\, of\,\,(\ref{eberhard})}\,\, \forall\,
\eta_a \,\, {\rm when} \,\, \eta_b=1~.
\end{equation}

Note that the conclusion (\ref{best})
of \emph{Case 3} is independent of the value of $\alpha$, i.e.,
the test can be applied to an infinity of Hardy states
given by Eq.~(\ref{hardy}). To have an idea of the minimum detection efficiencies 
required in a real test, let us first consider the case in which $\eta_b=0.90$
and $\alpha=0.9$ or $\alpha=\alpha_H$ (the value $\alpha_H=0.618$ corresponds 
to maximize Hardy's fraction (\ref{prima-2})):
$P^{\rm I}_{\rm QM}(a_+,a_+)|_{\alpha=0.9}=0.025$ and $\eta_a(\alpha=0.9)>0.25$,
while $P^{\rm I}_{\rm QM}(a_+,a_+)|_{\alpha=\alpha_H}=0.090$ and 
$\eta_a(\alpha=\alpha_H)>0.52$.

\begin{figure}[h]
\begin{center}
\mbox{\epsfig{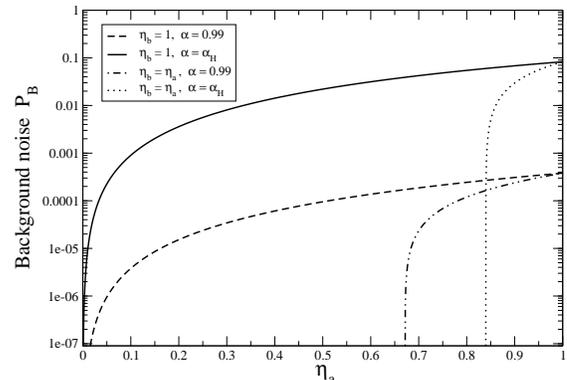}}
\caption{Maximum tolerable background noise
in a loophole--free experiment with inequality (\ref{eberhard})
as a function of $\eta_a$ for {\em Case 3} and: 
$\eta_b=1$ and $\alpha=0.99$; 
$\eta_b=1$ and $\alpha=\alpha_H=0.618$;
$\eta_b=\eta_a$ and $\alpha=0.99$;
$\eta_b=\eta_a$ and $\alpha=\alpha_H=0.618$.}
\label{figure}
\end{center}
\end{figure}

In real experiments, measurements are affected by noise
(i.e., by counts which do not originate from the entangled state under study)
in addition to inefficiencies in the detection. 
For white noise, represented by a background joint probability $P_B$
independent of the measurement settings,
the state subject to observation 
is not Hardy's state (\ref{hardy}) but rather the mixture:
\begin{equation}
\rho=(1-P_B)|\psi_H\rangle \langle \psi_H|+ P_B \frac{\one}{4}~.
\end{equation}
The ratio of Eq.~(\ref{viol1}) for {\em Case 3} thus becomes:
\begin{equation}
\label{viol2}
H_{\rm QM}=\frac{\displaystyle (1-P_B)\frac{
(1-\alpha^2)^2\alpha^2}{2-\alpha^2}\eta_a^2+\frac{\displaystyle P_B}{4}}
{\displaystyle 5\frac{\displaystyle P_B}{4}+(1-P_B)\frac{\displaystyle
(1-\alpha^2)^2}{2-\alpha^2}2\eta_a(1-\eta_b)}~,
\end{equation}
and inequality (\ref{eberhard}) is violated when,
for given values of $\eta_a$ and $\eta_b$, the background noise is
limited by:
\begin{equation}
\label{viol2a}
P_B\leq P^{\rm max}_{B}=\frac{\displaystyle
\frac{(1-\alpha^2)^2}{2-\alpha^2}\eta_a [\alpha^2 \eta_a-2(1-\eta_b)]}
{1+\displaystyle \frac{(1-\alpha^2)^2}{2-\alpha^2}\eta_a
[\alpha^2 \eta_a -2(1-\eta_b)]}~.
\end{equation}
In Figure~\ref{figure} we show the maximum tolerable background noise 
in a loophole--free experiment adopting 
Eberhard inequality (\ref{eberhard})
as a function of $\eta_a$ for four relevant cases.
The results for $\eta_b=\eta_a$ and $\alpha=0.99$ correspond
to a critical efficiency $\eta^{\rm min}_a=2/3\simeq 0.67$, in agreement
with what found in Ref.~\cite{Eb93}, while for
$\eta_b=\eta_a$ and $\alpha=\alpha_H$,  
$\eta^{\rm min}_a=4/(7-\sqrt{5})\simeq 0.84$. Instead, when $\eta_b=1$,
and independently of the value of $\alpha$,
any value of $\eta_a$ allows a loophole--free experiment
when the background is limited to the value given by Eq.~(\ref{viol2a}):
for instance, for $\alpha=\alpha_H$ 
a genuine violation of Eberhard inequality is possible
using $\eta_a>0.33$ for a background of 1\%.
It turns out that for values of $\eta_a$ and $\eta_b$ which allow a 
loophole--free test, the value of $\alpha$ which maximizes
the tolerable noise is always larger than $\alpha_H$:
in the limiting case of $\eta_a=\eta_b=1$,
$P^{\rm max}_B$ is maximum for $\alpha=\alpha_H$:
$P^{\rm max}_B(\eta_a=\eta_b=1, \alpha=\alpha_H)=(13-5\sqrt{5})/22\simeq
0.083$.

\begin{figure}[h]
\begin{center}
\mbox{\epsfig{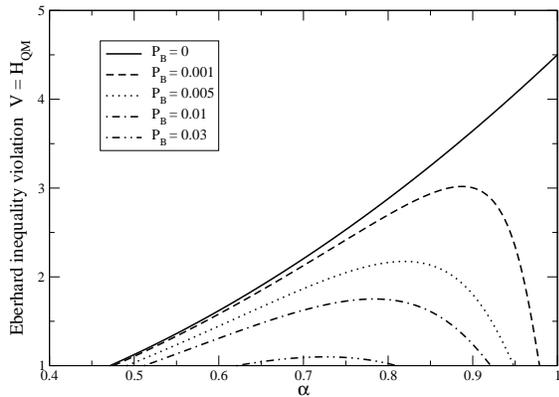}}
\caption{Violation $V=H_{\rm QM}$ of the
Eberhard inequality (\ref{eberhard}) predicted by quantum mechanics
for {\em Case 1} with $\eta=0.9$
as a function of the parameter $\alpha$ defining Hardy's state (\ref{hardy}).}
\label{violation}
\end{center}
\end{figure}

Another important question in a Bell--test is the
amount of violation predicted by quantum mechanics for the Bell inequality. 
For our Eberhard inequality the violation is given
by $V=H_{\rm QM}/H^{\rm max}_{\rm LR}=H_{\rm QM}$.
Considering for instance atom--atom entanglement \cite{Ro09}, for which
efficiencies as high as 90\% can be reached, for different
values of the background noise we obtain the results of $V$ vs $\alpha$
of Figure~\ref{violation}. For moderate noise
the expected violation can be large (to be compared, for instance,
with the maximum violation $(3+2\sqrt{2})/3\simeq 1.94$ obtained 
for the ideal case with maximally entangled states
and four measurement settings). We also note that
$V\to \infty$ for $\eta_b\to 1$ and $P_B\to 0$.

The other two Eberhard inequalities which can be derived from Hardy's
reasoning applied to Eqs.~(\ref{seconda}), (\ref{terza}), (\ref{quarta})
and (\ref{prima-2}) 
differ from the one in Eq.~(\ref{eberhard}) for the two joint probabilities
involving undetected events: one contains the sum $P(a_+,b_0)+P(b_0,b_+)$,
the other $P(b_+,b_0)+P(b_0,a_+)$. We do not discuss
these additional inequalities since,
for each one of the three special cases previously analyzed,
a loophole--free test with each one of them would require 
values for the detection efficiency thresholds
and the tolerable noise which are less convenient 
than obtained for inequality (\ref{eberhard}).

In conclusion, we have discussed a Bell--type test
involving a bipartite non--maximally entangled state
of the Hardy type and (unlike standard Bell--tests) 
the same pair of measurement bases for both parties.
As far as we know, the results we have obtained improve all previous discussions 
aimed at finding the bipartite entangled state and the Bell measurements bases
allowing one to refute local realism with the minimum possible
detection efficiencies.
In the design of new detection--loophole--free tests it is 
important to identify entangled systems for which one of the two 
required observables can be measured 
with very high efficiency: in the absence of noise,
a genuine Bell inequality violation is thus affordable even
with very low efficiencies for the other measurement.

The only system we know that enables observable asymmetric 
measurements consists of entangled neutral kaon pairs \cite{Br06}.
For kaons, lifetime measurements can be performed quite efficiently
($\eta_b\simeq 0.9$), but, unfortunately, strangeness measurements 
are still affected by very small efficiencies ($\eta_a< 0.01$).
On the contrary, for atom--atom entanglement
the proposed Bell--test allows large falsifications
of local realism (even greater than the well--known violations
predicted for maximally entangled states in ideal Bell--tests).


\end{multicols}
\end{document}